\documentclass[1p,dvips,numbers,sort]{elsarticle}
\usepackage{latexsym}
\usepackage{amsmath}
\usepackage{graphicx}
\usepackage{subfigure} 
\usepackage{here}

\bibpunct{[}{]}{,}{n}{}{;} 

\def\<#1>{\langle \hbox{}{#1}\hbox{} \rangle_{q}}
\def\trace{\mathrm{Tr}}
\def\G<#1>{\langle \hbox{}{#1}\hbox{} \rangle_{q}^{\mathrm{G}}}
\def\state#1{\left| #1 \right\rangle} 
\def\invstate#1{\left\langle #1 \right|} 
\def\tauopt{\tau_{\mathrm{opt}}}

\begin{document}
\begin{frontmatter}
\title{Application of optimization method to the $x^4$ model in the Tsallis nonextensive statistics}
\journal{}
\author{Masamichi Ishihara\corref{cor}}
\cortext[cor]{Corresponding author. Tel.: +81 24 932 4848; Fax: +81 24 933 6748.}
\ead{m\_isihar@koriyama-kgc.ac.jp}
\address{Department of Human Life Studies, Koriyama Women's University, Koriyama, 963-8503, Japan}
\begin{abstract}
We study the effects of the environment described by the Tsallis nonextensive statistics on physical 
quantities using an optimization method in the case of small deviation from the Boltzmann-Gibbs statistics. 
The $x^4$ model is used and the density operator is restricted to be a gaussian form. 
The variational parameter is the frequency $\Omega$ of a particle in the optimization method.
We obtain an approximate expression of free energy and of the expectation value of 
$\beta m \Omega^2 x^2 /2$, where $\beta$ is the inverse of the temperature and $m$ is the mass of a particle.
Numerically, the optimized frequency is estimated and 
the expectation value of $\beta m \Omega^2 x^2 /2$ is calculated.
The effects of the Tsallis nonextensive statistic for small deviation from the Boltzmann-Gibbs statistics are found:
1) the frequency modulation of a particle 
and 
2) the variation of the expectation value of $\beta m \Omega^2 x^2 /2$ at high temperature. 
\end{abstract}
\end{frontmatter}

\section{Introduction} 
An extended equilbrium statistics is often used to analyse phenomena.
Tsallis nonextensive statistics \cite{Tsallis} is an extended equilbrium statistics
and a parameter $q$ is introduced in this statistics.
This statistics may explain the phenomena that show power-law distribution and 
has been applied to various phenomena and methods,
such as particle distribution at high energies \cite{Wilk2009}
, network \cite{Hasegawa_physica} , simulated annealing algorithm \cite{Andricioaei_PRE53}, etc.
It is important to study the effects of the environment described by the Tsallis nonextensive statistics.

The $x^4$ model is basic and useful to study the effects of the environment.
The Hamiltonian $H(x)=C x^k$ was used to study power-law distribution \cite{Bashkirov2004}
and the effects of the anharmonic potential were studied using the $x^4$ model \cite{Ishihara2012} .
The $x^4$ model is also used to describe phase transitions.
Therefore, the $x^4$ model is a good base to study the effects of the environment describe by the Tsallis nonextensive statistics.

The optimization method is often used to estimate physical quantities.
An example to which an optimization method is applied is the Gaussian effective potential \cite{Stevenson1984,Stevenson1985}.  
The wave function is restricted to be a Gaussian form in the calculation of effective potential under the optimization method.
This method is extended to the calculation at finite temperature \cite{Hajj1988,Haugerud1991}.
The density operator is restricted with some parameters in an optimization method, 
and these parameters are determined by optimizing free energy. 
Many physical quantities can be calculated, because the density operator is determined. 
Therefore, optimizing the free energy with respect to the parameters is used to calculate physical quantities.

The absolute value $|1-q|$ is small in some systems \cite{Alberico2000,Biyajima2006,Osada2008,Wilk2009}, 
where this value is an index of the deviation from the Boltzmann-Gibbs statistics.
The $(1-q)$ expansion was often used in the previous studies \cite{Kohyama_Tsallis06,Ishihara2012}. 
The self-consistent equation for the energy is obtained generally in the Tsallis nonextensive statistics.
The $(1-q)$ expansion is useful to solve the equation order by order in $(1-q)$. 
Therefore, the $(1-q)$ expansion is used to solve the equation when the deviation from the Boltzmann-Gibbs statistics is small.

The purpose of this paper is to study the effects of the environment described by the Tsallis nonextensive statistics 
on physical quantities using an optimization method in the case of small $|1-q|$. 
The effects of the potential and of the environment are taken into the frequency of a free particle.
A parameter, frequency, is determined by optimizing the free energy in the optimization method.
The expectation value of $\beta m \Omega^2 x^2 / 2$ is also calculated when the parameter is given,
where $\beta$ is the inverse of the temperature $T$ and $m$ is the mass of a particle.
Therefore, we study the variation of the frequency and of the expectation value 
in the case of small $|1-q|$ using the optimization method. 

The effects of the Tsallis nonextensive statistics are found from the results:
1) the deviation from the Boltzmann-Gibbs statistics is observed by measuring the frequency modulation, 
and 
2) the effect of the statistics on the expectation value of $\beta m \Omega^2 x^2 / 2$ 
appears at high temperature.
Therefore, the deviation from the Boltzmann-Gibbs statistics is probably clarified. 

This paper is organized as follows. 
In section \ref{Sec:FreeEnergy}, 
we obtain the expression of the free energy with the restricted density operator in the $x^4$ model 
when the deviation from the Boltzmann-Gibbs statistics is small.
We also obtain the expression of the expectation value of $\beta m \Omega^2 x^2 / 2$.
In section \ref{Sec:NumericalCalc}, we calculate the frequency 
and the expectation value of $\beta m \Omega^2 x^2 / 2$ numerically.
Section \ref{Sec:Conclusion} is assigned for discussion and conclusion.

\section{Free energy of the $x^4$ model in the Tsallis nonextensive statistics with an optimization method}
\label{Sec:FreeEnergy}
\subsection{Tsallis nonextensive statistics and optimization method}
A parameter $q$ is introduced in the Tsallis nonextensive statistics. 
This statistics is equivalent to the Boltzmann-Gibbs statistics when $q$ is equal to $1$.

The density operator $\rho$ is defined by
\begin{equation}
\rho := \frac{1}{Z_q} \left[ 1 - (1-q) \frac{\beta}{c_q} \left( H - U_q \right) \right]^{1/(1-q)} 
,
\end{equation}
where $\beta$ is the inverse of the temperature $T$, $c_q$ is a coefficient, 
$U_q$ is the expectation value of Hamiltonian $H$, and $Z_q$ is the partition function.
The partition function $Z_q$ is defined by
\begin{equation}
Z_q :=  \trace \left[ 1 - (1-q) \frac{\beta}{c_q} \left( H - U_q \right)\right]^{1/(1-q)} .
\end{equation}
The quantities, $Z_q$ and $c_q$, are related to each other: $c_q = \left(Z_{q}\right)^{(1-q)}$ .
The expectation value of the physical quantity $O$ is defined by
\begin{equation}
\<O> := \frac{\trace \left[ \left( \rho^q \right) O \right]}{\trace \left(\rho^q \right) } .
\label{expectation_value}
\end{equation}
The entropy $S_q$ and the free energy $F_q$ are defined by
\begin{align}
S_q &:= \frac{1}{(1-q)} \left[ \trace \left( \rho^q \right) -1  \right] , \\
F_q &:= U_q - T S_q . 
\end{align}
The physical quantities are calculated with the free energy $F_q$. 

The calculation of the expectation value of a physical quantity is not always easy in the Tsallis nonextensive statistics,  
because the energy $U_q$ is included in the definition of the density operator. 
Therefore, we apply an optimization method to calculate the quantities approximately. 

We restrict the density operator $\rho$ to be a gaussian form. 
The Hamiltonian $H$ is replaced by the free Hamiltonian $H_0$ whose frequency is $\Omega$. 
That is, we use the free Hamiltonian: 
\begin{equation}
H_0(\Omega) :=  \frac{p^2}{2m} + \frac{1}{2} m \Omega^2 x^2 ,
\end{equation}
where $x$ is coordinate, $p$ is momentum, and $m$ is mass of a particle. 
The restricted density operator is denoted as $\rho^{\mathrm{G}}$: 
\begin{equation}
\rho^{\mathrm{G}} := \frac{1}{Z_q^{\mathrm{G}}} \left[ 1 - (1-q) \frac{\beta}{c_q^{\mathrm{G}}} \left( H_{0}(\Omega) - U_q^{\mathrm{G}} \right) \right]^{1/(1-q)} ,
\label{eqn:definition:rho_G}
\end{equation}
where $c_q^{\mathrm{G}}$ is a constant, $Z_q^{\mathrm G}$ is the partition function. 
The relation $c_q^{\mathrm{G}} = \left(Z_{q}^{\mathrm{G}} \right)^{(1-q)}$ is imposed in Eq.~\eqref{eqn:definition:rho_G}.  
The quantity $U_q^{\mathrm{G}}$ is the expectation value of the Hamiltonian 
when the density operator is $\rho^{\mathrm{G}}$.
Therefore, the self-consistent equation of the energy remains:
\begin{equation}
U_q^{G} = \G<H> = \frac{\trace \left[ \left(\rho^{\mathrm G} \right)^q  H \right]}{\trace \left[\left( \rho^{\mathrm G} \right)^q \right] } .
\label{eqn:self-consistent}
\end{equation}
The free energy is also a function of $\Omega$, 
and we denote this free energy as $F_{q}^{\mathrm{G}}$.

The equation to determine $\Omega$ will be obtained by optimizing $F_q^{\mathrm{G}}$ with respect to the parameter $\Omega$. 
The energy $U_q^G$ will be determined using the self-consistent equation and the optimization condition.

\subsection{Application to the $x^4$ model}
\label{Subsec:Application}
In this subsection, the optimization method is applied to the $x^4$ model. 
The deviation $\epsilon$ from $q=1$ is defined by $\epsilon = 1 - q$. 
We use the parameter $\epsilon$ instead of $q$.
In the present paper, we deal with the case that the absolute value of $\epsilon$ is small. 

The Hamiltonian of the $x^4$ model is given by
\begin{equation}
H = \frac{p^2}{2m} + \frac{1}{2} m \omega^2 x^2 + \lambda x^4 .
\end{equation}
The eigenstate of the free Hamiltonian $H_0(\Omega)$ is introduced: 
\begin{equation}
H_0(\Omega) \state{n} = \hbar \Omega \left( n+ \frac{1}{2} \right) \state{n} . 
\end{equation}
The Hamiltonian is divided into two parts: $H = H_0(\Omega) + H_{\mathrm{int}}$. 
The expectation value of $H_{\mathrm{int}}$ with respect to $\state{n}$ is 
\begin{subequations}
\begin{align}
& \invstate{n} H_{\mathrm{int}} \state{n} = A n^2 + 2B n + B , \\ 
& A = \left(\frac{3\lambda \hbar^2}{2m^2\Omega^2} \right), \quad 
  B = \left( \frac{3\lambda \hbar^2}{4m^2\Omega^2} \right) + \left( \frac{\hbar \omega}{4} \right) \left( \frac{\omega}{\Omega} - \frac{\Omega}{\omega} \right)
. 
\label{def:A:B}
\end{align}
\end{subequations}

First, we attempt to calculate the entropy $S_{q=1-\epsilon}$ to the order $\epsilon$. 
For simplicity, we define $\tilde{U}_{q}^{\mathrm{G}}$ by 
\begin{equation}
\tilde{U}_{q}^{\mathrm{G}} = U_q^{\mathrm{G}} - \frac{1}{2} \hbar \Omega. 
\end{equation}
The energy $\tilde{U}_{q=1-\epsilon}^{\mathrm{G}}$, the coefficient $c_q^{\mathrm{G}}$, and the partition function $Z_{q}^{\mathrm{G}}$ 
are expanded as series of $\epsilon$:
\begin{subequations}
\begin{align}
& \tilde{U}_{q=1-\epsilon}^{\mathrm{G}} = \tilde{U}_{0} - \epsilon \tilde{U}_{1} + O\left( \epsilon^2 \right), 
\label{eqn:energy:expansion}\\
& c_{q=1-\epsilon}^{\mathrm{G}} = c_0 - \epsilon c_1 + O\left( \epsilon^2 \right), 
\label{eqn:c:expansion}\\
& Z_{q=1-\epsilon}^{\mathrm{G}} = Z_0 - \epsilon Z_1 + O\left( \epsilon^2 \right). 
\label{eqn:Zq:expansion}
\end{align}
\end{subequations}
We omit the index $\mathrm{G}$ for simplicity 
in the right hand sides of Eqs.\eqref{eqn:energy:expansion}, \eqref{eqn:c:expansion} and \eqref{eqn:Zq:expansion}.
Moreover, we introduce the following functions:
\begin{subequations}
\begin{align}
K^{(j)} &:= \sum_{n=0}^{\infty} n^j \exp \left( - \beta_q \hbar \Omega n \right) , \qquad \beta_{q} = \beta / c_q^{\mathrm{G}} , 
\label{def:Kj}\\
\tilde{K}^{(j)} &:= K^{(j)} / K^{(0)} .
\end{align}
\end{subequations}
The expression of the function $K^{(j)}$ is given in \ref{app:Kj}.
These functions $K^{(j)}$ and $\tilde{K}^{(j)}$ depend on $\epsilon$. 
Therefore, the functions are expanded as series of $\epsilon$ as follows:
\begin{subequations}
\begin{align}
K^{(j)} &= K_0^{(j)} - \epsilon K_1^{(j)} + O\left(\epsilon^2\right) , \\
\tilde{K}^{(j)} &= \tilde{K}_0^{(j)} - \epsilon \tilde{K}_1^{(j)} + O\left(\epsilon^2\right) . 
\end{align}
\end{subequations}
The coefficients $c_{0}$ and $c_{1}$ with these functions are given by
\begin{subequations}
\begin{align}
c_0 &= 1 ,\\
c_1 &= - \ln Z_{0} = -\frac{\beta \tilde{U}_0}{c_0} - \ln K_0^{(0)} .  
\end{align}
\end{subequations}
The entropy $S_{q=1-\epsilon}$ can be expressed as follows:
\begin{subequations} 
\begin{align}
S_{q=1-\epsilon} &= S_0 - \epsilon S_1 +O\left(\epsilon^2\right) ,\\
S_0 &= \ln K_0^{(0)} + \tilde{K}_{0}^{(1)} \left( \frac{\beta \hbar \Omega}{c_0} \right) , \\
S_1 &= \frac{1}{2} \left( \tilde{K}_{0}^{(3)} - \tilde{K}_{0}^{(1)} \tilde{K}_{0}^{(2)} \right) \left( \frac{\beta \hbar \Omega}{c_0} \right)^3  
    \nonumber \\ & \qquad 
      + \left\{ \left(\frac{\beta \tilde{U}_0}{c_0} \right) \left( \tilde{K}_{0}^{(1)} \right)^2 
        - \left[ \frac{1}{2} + \left(\frac{\beta \tilde{U}_{0}}{c_0} \right) \right] \tilde{K}_{0}^{(2)} \right\} \left( \frac{\beta \hbar \Omega}{c_0} \right)^2 
    \nonumber \\ & \qquad 
      + \left[ \tilde{K}_{1}^{(1)} - \left( \frac{c_1}{c_0} \right) \tilde{K}_{0}^{(1)} - \tilde{K}_{0}^{(1)} \ln K_{0}^{(0)} \right] \left( \frac{\beta \hbar \Omega}{c_0} \right)
      + \left[ \frac{K_1^{(0)}}{K_0^{(0)}} - \frac{1}{2} \left(\ln K_0^{(0)} \right)^2 \right] .
\end{align}
\end{subequations} 

Next, we calculate the energy $\tilde{U}_{q=1-\epsilon}$. 
The function $I_{j}^{(q)}$ is defined for simplicity and is given in \ref{def:Ijq}. 
The function  $I_{j}^{q}$ is expanded as series of $\epsilon$: 
\begin{equation}
I_{j}^{q=1-\epsilon} = I_{j}^{(0)} - \epsilon I_{j}^{(1)} + O(\epsilon^2) . 
\end{equation}
The functions $\tilde{U}_{0}$ and $\tilde{U}_{1}$ in Eq.~\eqref{eqn:energy:expansion} are given to the order $\epsilon$: 
\begin{subequations}
\begin{align}
\tilde{U}_{0} &= \sum_{j=0}^{4} I_{j}^{(0)} \tilde{K}_{0}^{(j)} ,\\
\tilde{U}_{1} &= \sum_{j=0}^{4} I_{j}^{(1)} \tilde{K}_{0}^{(j)} + \sum_{j=0}^{4} I_{j}^{(0)} \tilde{K}_{1}^{(j)} 
\nonumber \\ & \quad 
              - \left[\left(\frac{\beta \tilde{U}_0}{c_0} \right)+ \ln K_0^{(0)} 
                + \frac{1}{2} \left( \frac{\beta \hbar \Omega}{c_0} \right)^2 \tilde{K}_{0}^{(2)}
                \right. \nonumber \\ & \qquad \left. 
                - \left( \frac{\beta \hbar \Omega}{c_0} \right) \left( \frac{\beta \tilde{U}_0}{c_0} \right) \tilde{K}_{0}^{(1)}
                + \frac{1}{2} \left( \frac{\beta \tilde{U}_0}{c_0} \right)^2 \tilde{K}_{0}^{(0)}
                \right] 
                \left[ \sum_{j=0}^{4} I_{j}^{(0)} \tilde{K}_{0}^{(j)} \right] . 
\end{align}
\end{subequations}

We can obtain the free energy $F_q^{\mathrm{G}} = U_{q}^{\mathrm{G}} - T S_{q}^{\mathrm{G}}$ to the order $\epsilon$ 
with the expression of $\tilde{U}_q^{\mathrm{G}}$ and of $\tilde{S}_q^{\mathrm{G}}$. 

The fluctuation is given by the expectation value $\G<x^2>$.
We obtain $\beta m \Omega^2 \G<x^2> / 2 $ instead of $\G<x^2>$. 
\begin{align}
& 
\frac{1}{2} \beta m \Omega^2 \G<x^2>
\nonumber \\ & 
= 
\frac{1}{2} (\beta \hbar \Omega) 
\left\{
\frac{1}{2} + \tilde{K}_{0}^{(1)}
\right. \nonumber \\ & \qquad \left.
- \epsilon  
\left[
\tilde{K}_{1}^{(1)} 
+ \frac{1}{2} (\beta \hbar \Omega)^2 \left( \tilde{K}_0^{(3)} - \tilde{K}_0^{(2)} \tilde{K}_0^{(1)} \right)  
+ (\beta \hbar \Omega) \left( 1 + \beta \tilde{U}_0 \right) 
  \left( \left( \tilde{K}_0^{(1)} \right)^2 - \tilde{K}_0^{(2)} \right) 
\right]
\right\}
.
\label{eqn:flucutuation:for_numerical}
\end{align}
We calculate the value given by Eq.~\eqref{eqn:flucutuation:for_numerical} numerically in the following section.

\section{Numerical Calculations}
\label{Sec:NumericalCalc}
In the numerical calculations, 
we introduce dimensionless parameters: $\gamma=\beta \hbar \omega$, $\Lambda = (\lambda \hbar)/(m^2 \omega^3)$ 
and $\tau=\Omega/\omega$.
We attempt to obtain the value of $\tau$ at the minimum of $\beta F_q^{\mathrm{G}}$ as a function of $\tau$. 
This value is represented as $\tau_{\rm{opt}}$.
We also attempt to calculate the quantity $\beta m \Omega^2 \G<x^2> /2$.

\begin{figure}[H]
\subfigure[$\epsilon=0$]{
\includegraphics[width=0.45\textwidth]{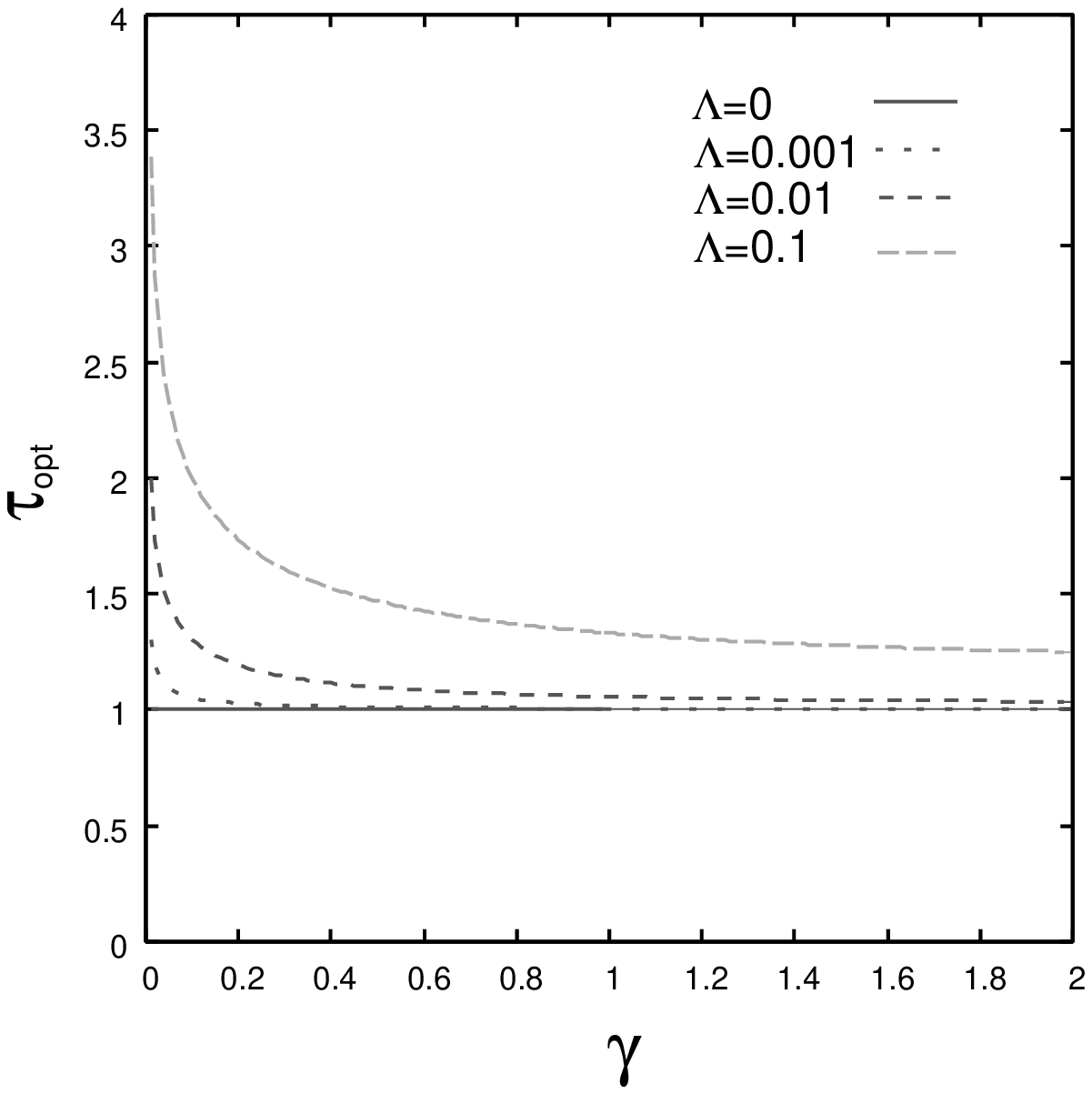}
\label{fig:gamma-Lambda-dep:epsilon=0}
}
\subfigure[$\epsilon=0.1$]{
\includegraphics[width=0.45\textwidth]{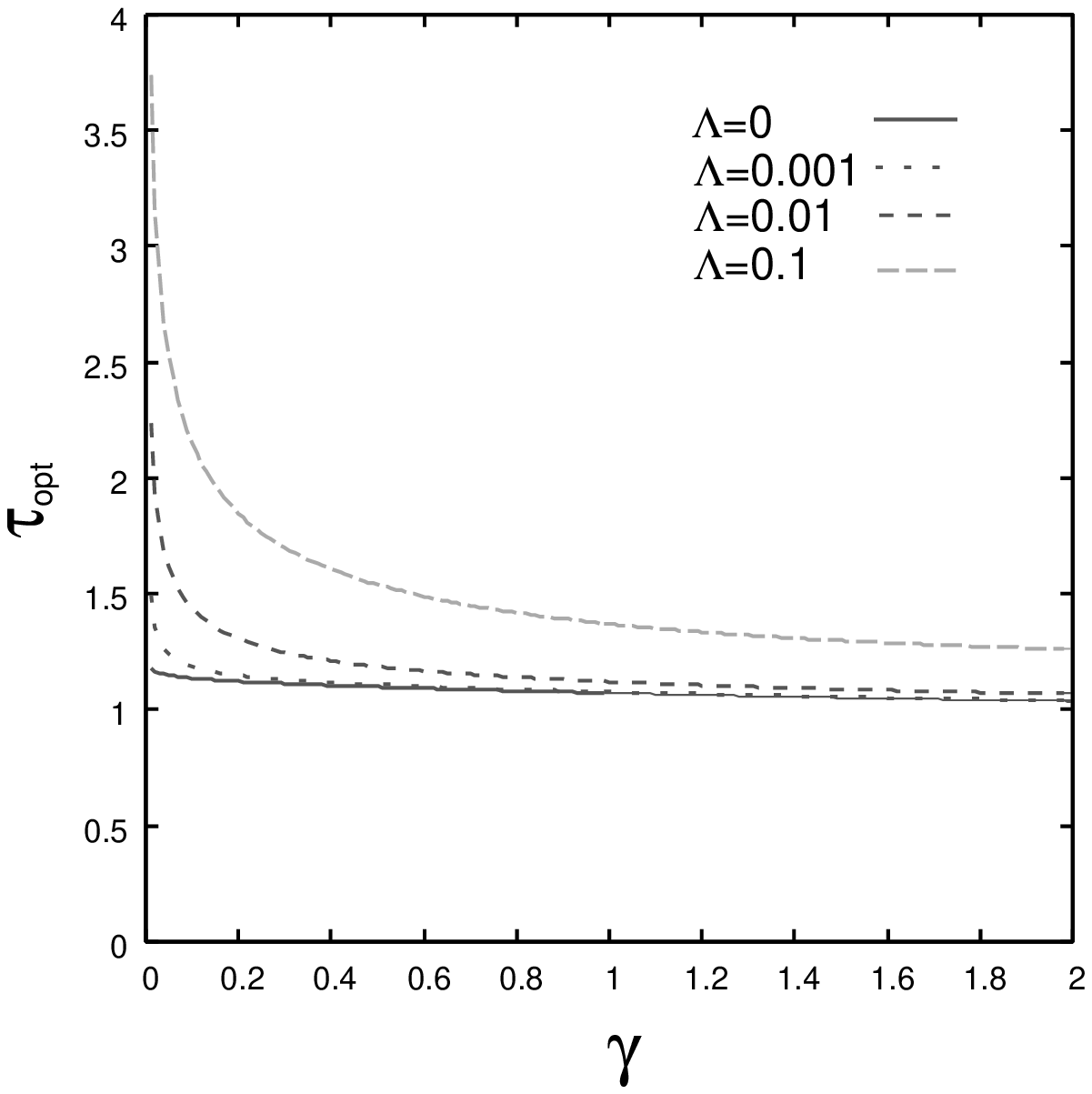}
\label{fig:gamma-Lambda-dep:epsilon=0.1}
}
\caption{The $\gamma$ dependence of $\tauopt$ for various $\Lambda$}
\label{fig:gamma-Lambda-dep}
\end{figure}

Figure~\ref{fig:gamma-Lambda-dep} shows the $\gamma$ dependence of $\tauopt$ for various $\Lambda$.
Figure~\ref{fig:gamma-Lambda-dep:epsilon=0} shows the $\gamma$ dependence at $\epsilon=0$ (the Boltzmann-Gibbs statistics). 
The value $\tauopt$ is just 1 when the $x^4$ term is absent.
The value $\tauopt$ is large for small $\gamma$ and for large $\Lambda$ in the figure.
These results indicate that the shift of frequency is large at high temperature and for strong interaction. 
Figure~\ref{fig:gamma-Lambda-dep:epsilon=0.1} shows the $\gamma$ dependence at $\epsilon=0.1$. 
The global behavior of $\tauopt$ in Figure~\ref{fig:gamma-Lambda-dep:epsilon=0.1} is similar 
to that in Figure~\ref{fig:gamma-Lambda-dep:epsilon=0}.
The value $\tauopt$ for $\epsilon=0.1$ is larger than that for $\epsilon=0$. 
The difference between the Boltzmann-Gibbs and Tsallis nonextensive statistics is reflected in frequency. 

\begin{figure}[H]
\subfigure[$\Lambda=0$]{
\includegraphics[width=0.45\textwidth]{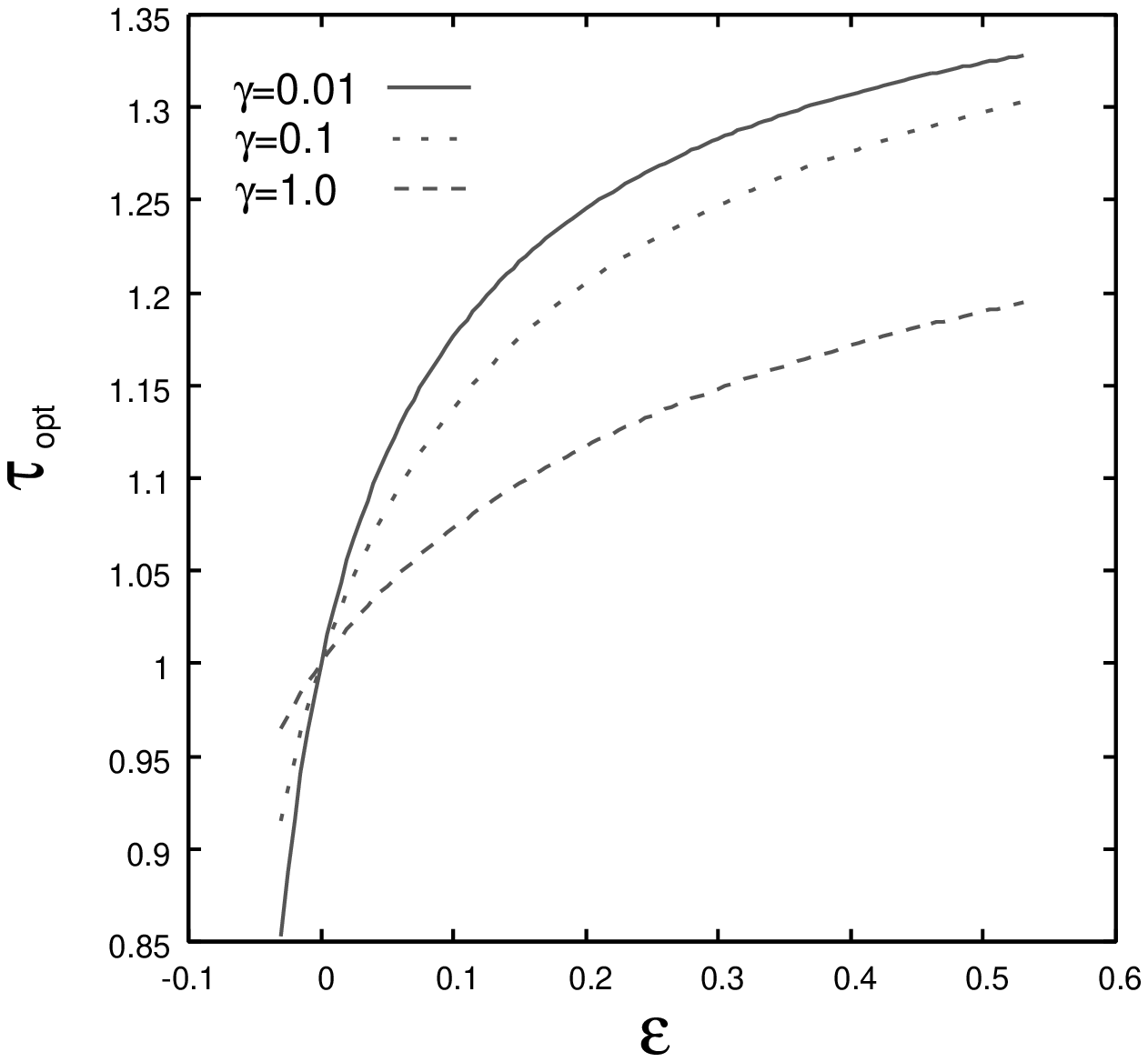}
\label{fig:epsilon_dep:Lambda0.000}
}
\subfigure[$\Lambda=0.01$]{
\includegraphics[width=0.45\textwidth]{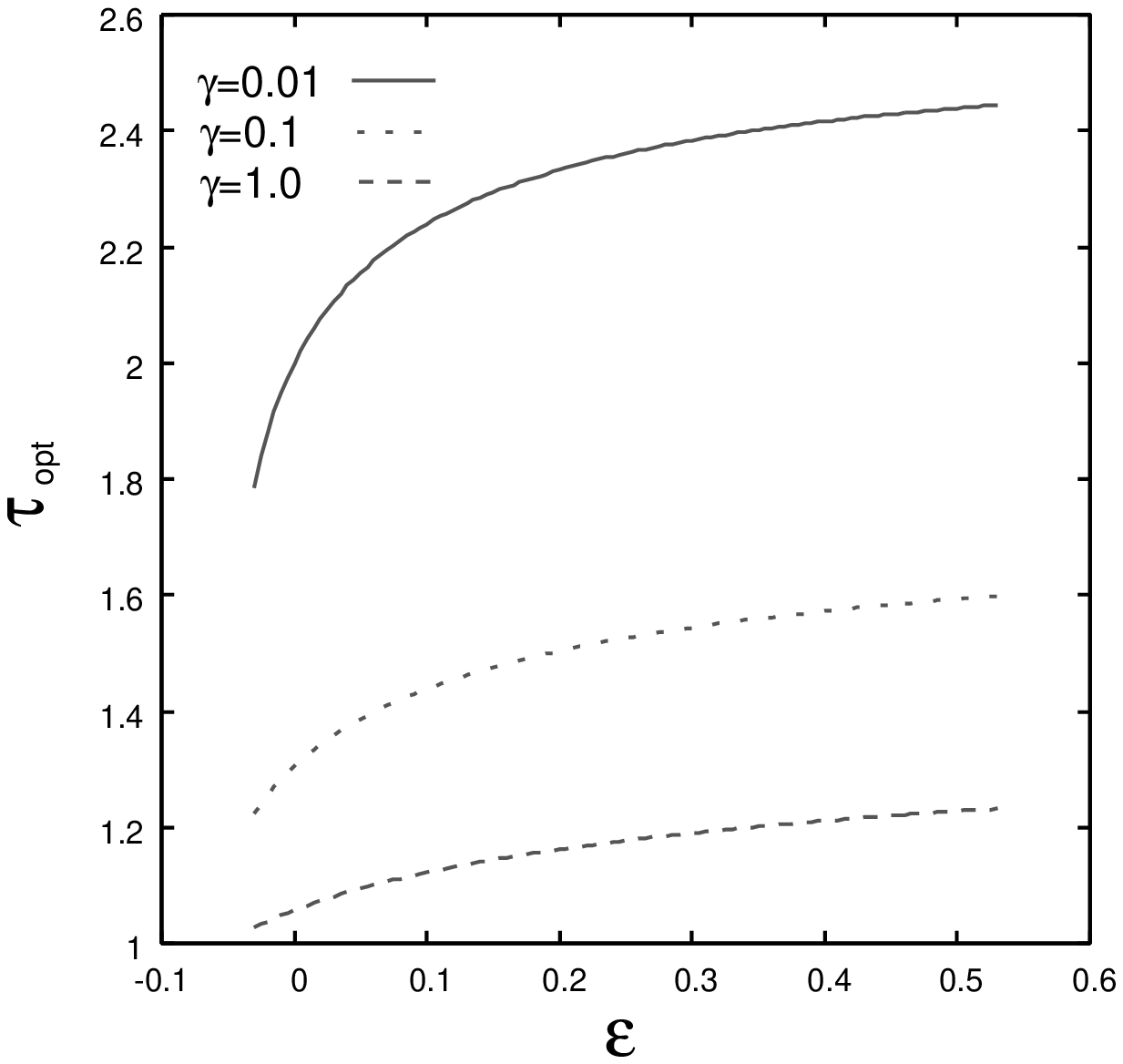}
\label{fig:epsilon_dep:Lambda0.010}
}
\caption{The $\epsilon$ dependence of $\tauopt$ for various $\gamma$}
\label{fig:epsilon_dep}
\end{figure}

Figure~\ref{fig:epsilon_dep} shows the $\epsilon$ dependence of $\tauopt$ for various $\gamma$. 
Figure~\ref{fig:epsilon_dep:Lambda0.000} shows the  $\epsilon$ dependence at $\Lambda=0$.
The curves intersect at a point: $\epsilon=0$ and $\tauopt=1$, 
because the Boltzmann-Gibbs statistics corresponds to $\epsilon=0$.
The deviation from $\tauopt=1$ is large for small $\gamma$ in Figure~\ref{fig:epsilon_dep:Lambda0.000}, 
as shown in Figure~\ref{fig:gamma-Lambda-dep}.
Figure~\ref{fig:epsilon_dep:Lambda0.010} shows the $\epsilon$ dependence at $\Lambda=0.01$.
The $\epsilon$ dependence for $\Lambda=0.01$ is similar to that for $\Lambda=0$,
while the value $\tauopt$ for $\Lambda=0.01$ is larger than that for $\Lambda=0$ at the same $\epsilon$, because of the interaction.  

\begin{figure}[H]
\centering
\subfigure[$\epsilon=-0.02$]{
\includegraphics[width=0.45\textwidth]{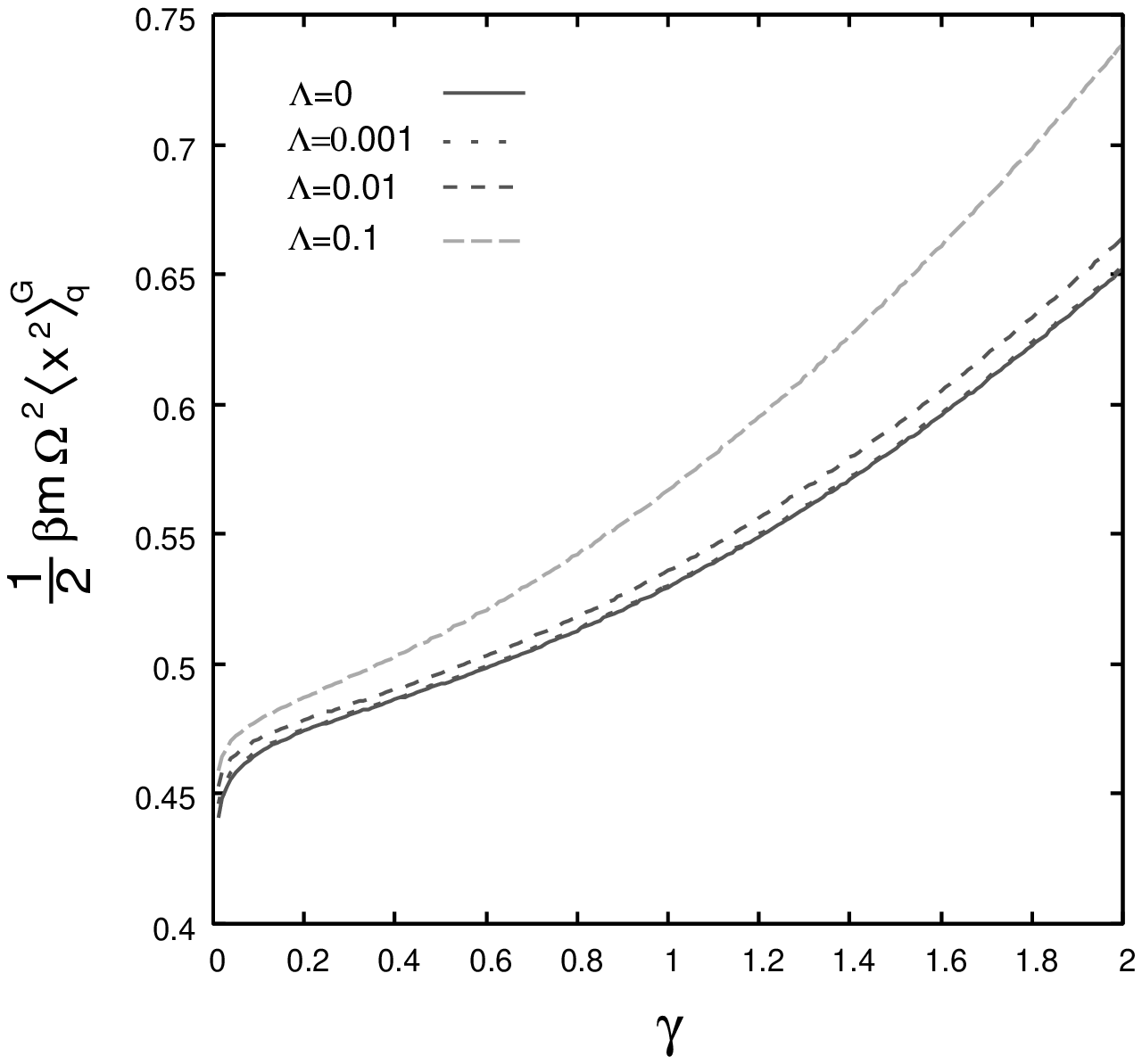}
\label{fig:x2:gamma-Lambda-dep:epsilon=-0.02}
}
\subfigure[$\epsilon=0$]{
\includegraphics[width=0.45\textwidth]{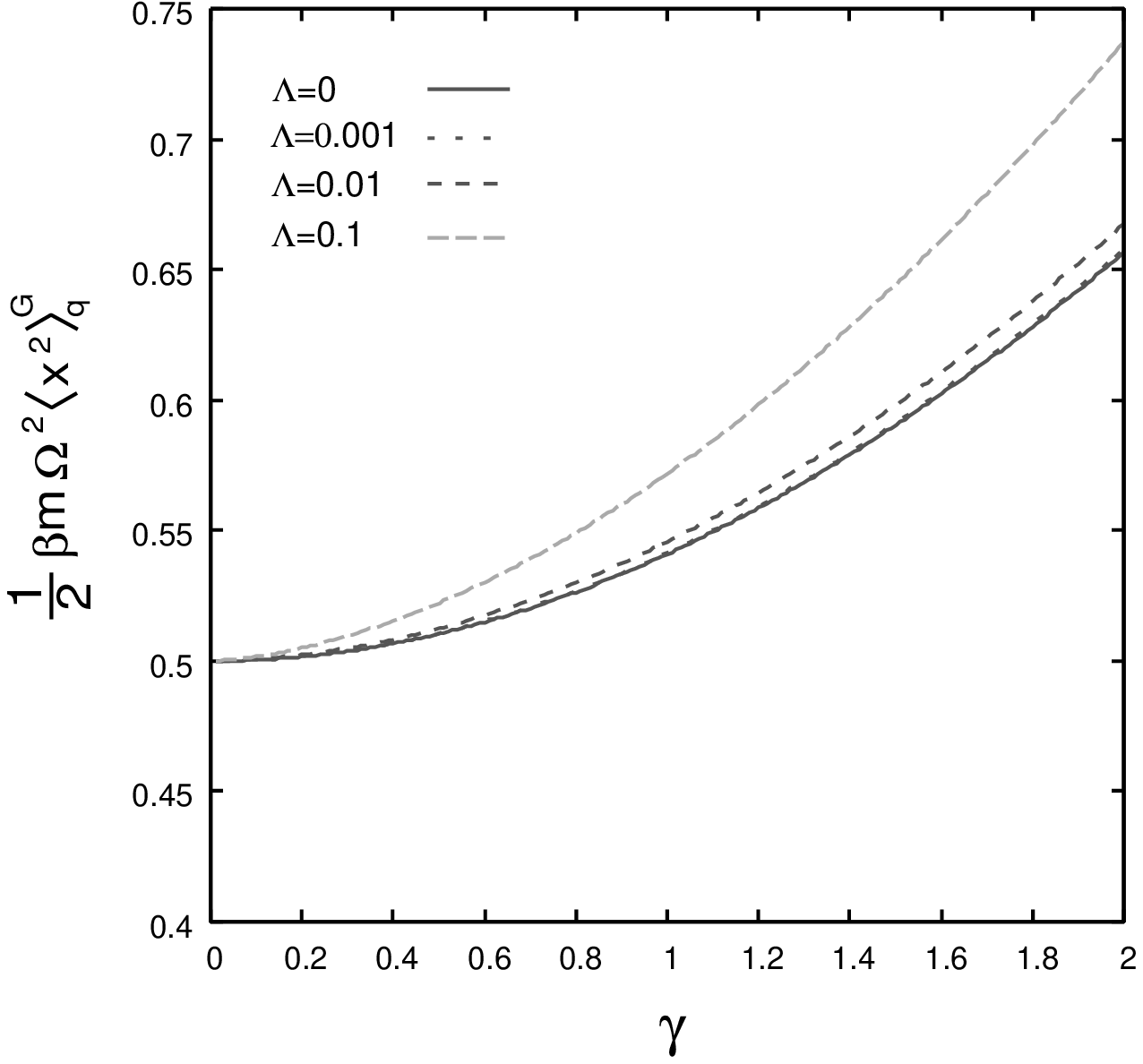}
\label{fig:x2:gamma-Lambda-dep:epsilon=0.00}
}
\\
\subfigure[$\epsilon=0.02$]{
\includegraphics[width=0.45\textwidth]{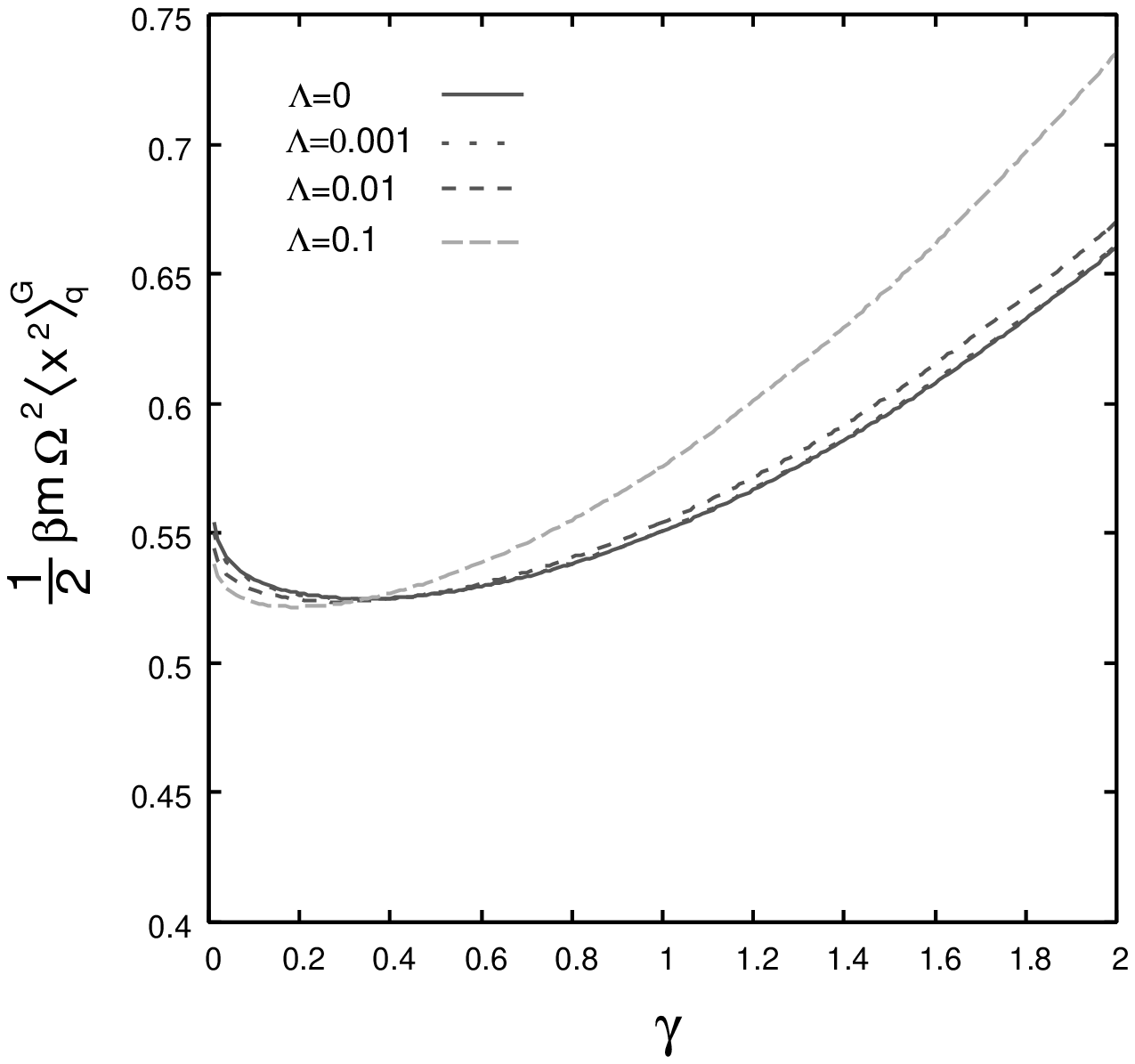}
\label{fig:x2:gamma-Lambda-dep:epsilon=0.02}
}
\caption{The $\gamma$ dependence of $\beta m \Omega^2 \G<x^2> / 2$ for various $\Lambda$}
\label{fig:x2:gamma-Lambda-dep}
\end{figure}

The expectation value $\beta m \Omega^2 \G<x^2> / 2$ is estimated when $\tauopt$ is given.
Figure~\ref{fig:x2:gamma-Lambda-dep} shows 
the $\gamma$ dependence of the value $\beta m \Omega^2 \G<x^2> /2$ for various $\Lambda$. 
Figure~\ref{fig:x2:gamma-Lambda-dep:epsilon=-0.02} is the graph for $\epsilon=-0.02$, 
Figure~\ref{fig:x2:gamma-Lambda-dep:epsilon=0.00} is for $\epsilon=0$, 
and 
Figure~\ref{fig:x2:gamma-Lambda-dep:epsilon=0.02} is for $\epsilon=0.02$.
This value converges to 0.5 as $\gamma$ approaches zero for $\epsilon=0$. 
The effects of the Tsallis nonextensive statistics are shown 
in Figure~\ref{fig:x2:gamma-Lambda-dep:epsilon=-0.02} and in Figure~\ref{fig:x2:gamma-Lambda-dep:epsilon=0.02}.
The value $\beta m \Omega^2 \G<x^2> /2$ as a function of $\gamma$ has a minimum at $\gamma \neq 0$ for $\epsilon = 0.02$.
In contrast, the value for $\epsilon=-0.02$ is smaller than that for $\epsilon=0$. 
The value $\beta m \Omega^2 \G<x^2> /2$ increases with $\Lambda$ for almost all the value $\gamma$.
For small $\gamma$ in Figure~\ref{fig:x2:gamma-Lambda-dep:epsilon=0.02},
the value for small $\Lambda$ is larger than that for large $\Lambda$.

\begin{figure}[H]
\subfigure[$\Lambda=0$]{
\includegraphics[width=0.45\textwidth]{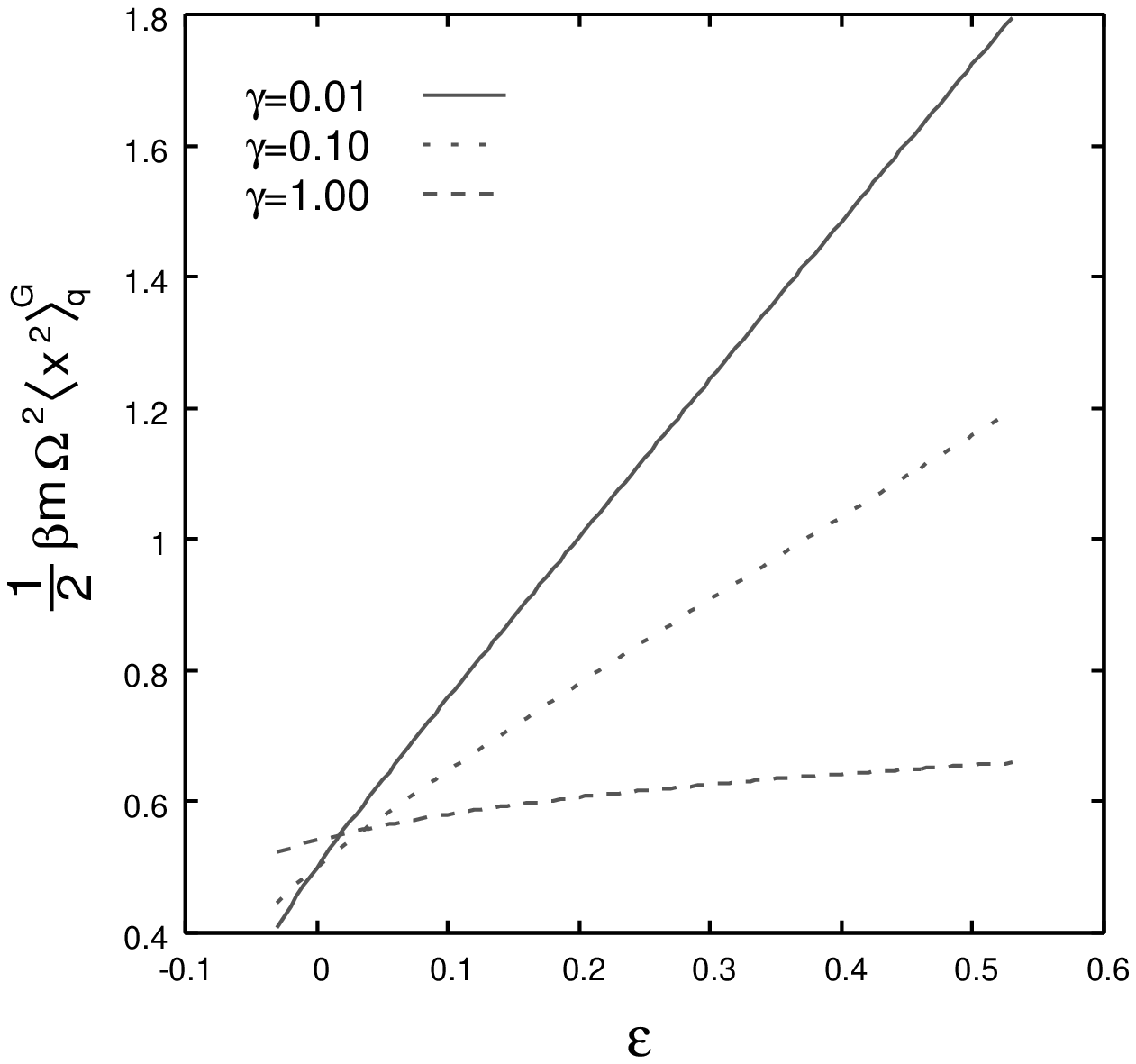}
\label{fig:x2:eps:Lambda0.000}
}
\subfigure[$\Lambda=0.01$]{
\includegraphics[width=0.45\textwidth]{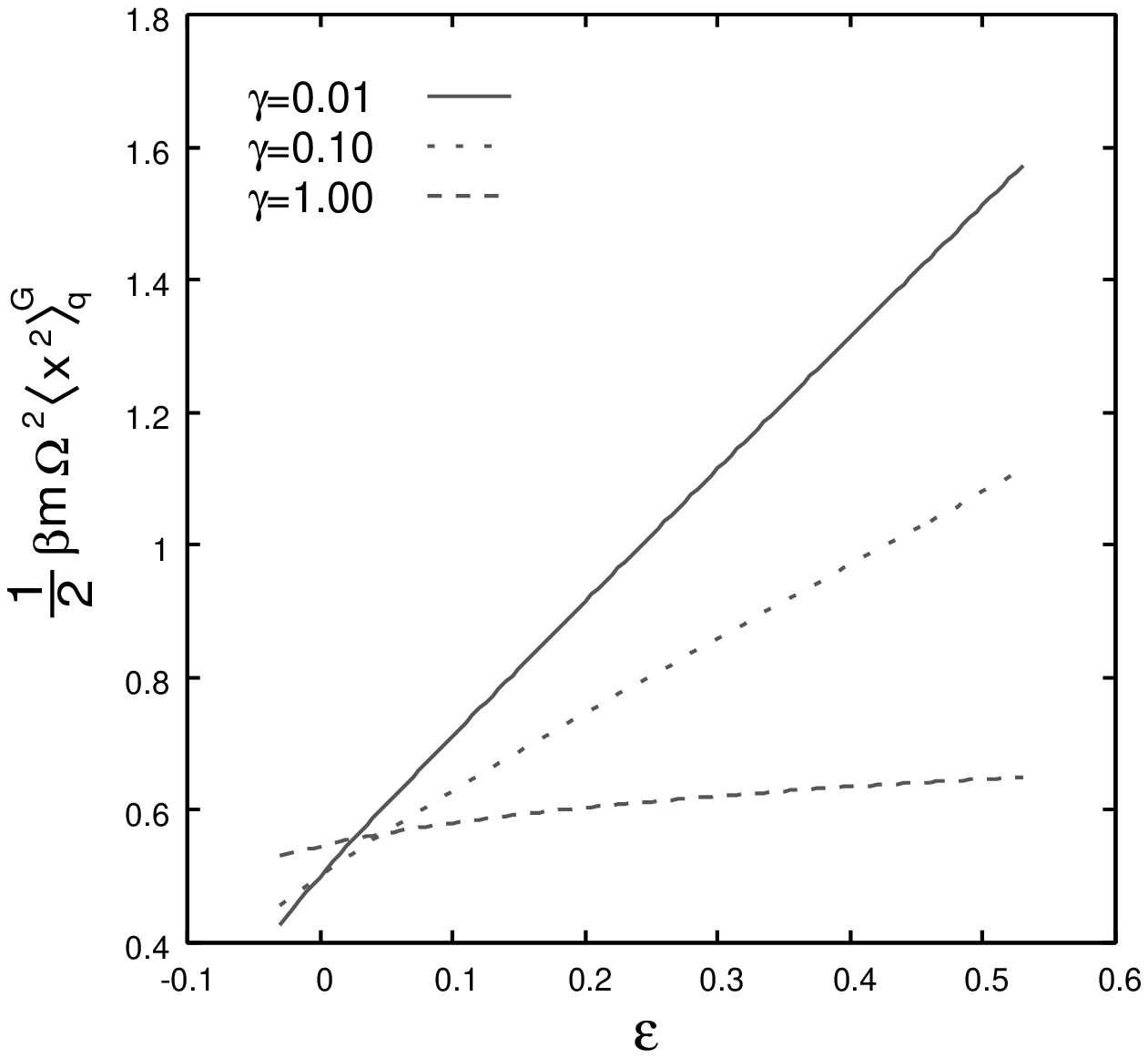}
\label{fig:x2:eps:Lambda0.010}
}
\caption{The $\epsilon$ dependence of $\beta m \Omega^2 \G<x^2>/2$ for various $\gamma$}
\label{fig:x2:eps:Lambda}
\end{figure}

Figure~\ref{fig:x2:eps:Lambda} shows the $\epsilon$ dependence of 
the value $\beta m \Omega^2 \G<x^2> /2$ for various $\gamma$. 
Figure~\ref{fig:x2:eps:Lambda0.000} is for $\Lambda=0$ and 
Figure~\ref{fig:x2:eps:Lambda0.010} is for $\Lambda=0.01$.
The $\epsilon$ dependence of the value for $\Lambda=0.01$ is similar to that for $\Lambda=0$.
The inclination of a curve is large for small $\gamma$. 
This fact indicates that the difference between the Boltzmann-Gibbs and Tsallis nonextensive statistics is apparent for small $\gamma$.

\section{Discussion and Conclusion}
\label{Sec:Conclusion}
In this paper, we studied the effects of the environment described by the Tsallis nonextensive statistics 
on physical quantities using an optimization method in the case of small $|1-q|$,
where $(1-q)$ is an index of the deviation from the Boltzmann-Gibbs statistics. 
We restricted the Hamiltonian in the density operator 
by the Hamiltonian of a free particle with the frequency $\Omega$,
where $\Omega$ is the variational parameter of the optimization method.
The $x^4$ model were used and the physical quantities were expanded as the series of $\epsilon=1-q$ to solve the equations order by order. 
We derived an approximate expression of the free energy and 
of the expectation value $\beta m \Omega^2 \G<x^2> /2$ in the optimization method.
We used the parameter $\tau$ that is defined 
by the ratio of the frequency $\Omega$ to the frequency $\omega$,
where $\omega$ is the frequency in the $x^4$ model. 
We obtained numerically the value of $\tau$ at the minimum of the free energy.
This value is represented as $\tau_{\mathrm{opt}}$.
The expectation value $\beta m \Omega^2 \G<x^2> /2$ was also calculated numerically with the value of $\tau_{\mathrm{opt}}$.

The results show that the optimized frequency $\Omega$ increases with the temperature. 
The frequency modulation is large at high temperature. 
The frequency modulation is large even when the deviation from the Boltzmann-Gibbs statistics is small.
Therefore, the deviation from the Boltzmann-Gibbs statistics is probably observed by measuring the frequency modulation. 
The effect of the statistics on the value $\beta m \Omega^2 \G<x^2> /2$ appears at high temperature.
The deviation from the Boltzmann-Gibbs statistics is also clarified by the temperature dependence of the value $\beta m \Omega^2 \G<x^2> /2$.
The existence of the minimum of the value $\beta m \Omega^2 \G<x^2> /2$ 
as a function of the inverse temperature $\gamma$ is a signal of the deviation from the Boltzmann-Gibbs statistics. 
The existence of the minimum implies that the parameter $q$ is smaller than 1. 
In contrast, the parameter $q$ is larger than 1 when the value $\beta m \Omega^2 \G<x^2> /2$ drops near the origin of $\gamma$. 
The value $\beta m \Omega^2 \G<x^2> /2$ for $q \neq 1$ differs from that for $q = 1 $ at high temperature.

It is well-known that the interaction modifies the frequency in both the Boltzmann-Gibbs and Tsallis nonextensive statistics. 
The $\epsilon$ dependence of the frequency modulation with interaction is similar to that without interaction. 
The frequency modulation with interaction is large, compared with that without interaction, 
in both the Boltzmann-Gibbs and Tsallis nonextensive statistics. 

In conclusion, 
the effects of the Tsallis nonextensive statistic for small deviation from the Boltzmann-Gibbs statistics 
in the $x^4$ model are found:
1) the frequency modulation of a particle and 
2) the variation of the expectation value of $\beta m \Omega^2 x^2 /2$ at high temperature.

We hope that this work is helpful for understanding the nonextensive statistics.

\makeatletter \renewcommand{\@biblabel}[1]{[#1]} \makeatother

\appendix
\section{Function $K^{(j)}$}
\label{app:Kj}
The function $K^{(j)}$ is defined by Eq.~\eqref{def:Kj}.
The expression of $K^{(j)}$ ($j=0,1,2,3,4)$ is given as follows:
\begin{subequations}
\begin{align}
K^{(0)} & = \frac{\exp(\beta_q \hbar \Omega)}{\left( \exp(\beta_q \hbar \Omega)  - 1 \right)} 
,\\
K^{(1)} & = \frac{\exp(\beta_q \hbar \Omega)}{\left( \exp(\beta_q \hbar \Omega)  - 1 \right)^{2}} 
,\\
K^{(2)} & = \frac{\left(\exp(2 \beta_q \hbar \Omega) + \exp(\beta_q \hbar \Omega) \right)}{\left( \exp(\beta_q \hbar \Omega)  - 1 \right)^{3}} 
,\\
K^{(3)} & = \frac{\left(\exp(3 \beta_q \hbar \Omega) + 4 \exp(2 \beta_q \hbar \Omega) + \exp(\beta_q \hbar \Omega) \right)}{\left( \exp(\beta_q \hbar \Omega)  - 1 \right)^{4}} 
,\\
K^{(4)} & = \frac{\left(\exp(4 \beta_q \hbar \Omega) + 11 \exp(3 \beta_q \hbar \Omega) + 11 \exp(2 \beta_q \hbar \Omega) + \exp(\beta_q \hbar \Omega) \right)}{\left( \exp(\beta_q \hbar \Omega)  - 1 \right)^{5}}
.
\end{align}
\end{subequations}
Therefore, the expression of the function $\tilde{K}^{(j)}$ is obtained from the above expressions of $K^{(j)}$.
Moreover, the functions $K^{(j)}_0$, $K^{(j)}_1$, $\tilde{K}^{(j)}_0$, and $\tilde{K}^{(j)}_1$  are obtained directly.

\section{Function $I_j^{q}$}
\label{def:Ijq}
For simplicity, the function $I_j^{q} (j = 0, 1, 2, 3, 4) $ is defined as follows:
\begin{subequations}
\begin{align}
& I_{4}^{q} := \epsilon \left[ - \frac{1}{2} \left(\beta_q \hbar \Omega \right)^2 A \right] ,\\
& I_{3}^{q} := \epsilon \left[ \left(1 + \beta_q \tilde{U}_{q}^{\mathrm{G}}\right) A 
    - \frac{1}{2} \left(\beta_q \hbar \Omega \right)  \left(2B + \hbar \Omega \right) \right] \left(\beta_q \hbar \Omega \right) , \\
& I_{2}^{q} := A - \epsilon \left[ \left(\beta_q \tilde{U}_{q}^{\mathrm{G}}\right) \left( 1 + \frac{1}{2} \beta_q \tilde{U}_q^{\mathrm{G}}\right) A 
    - \left(\beta_q \hbar \Omega \right) \left(1 + \beta_q \tilde{U}_q^{\mathrm{G}}\right)  \left(2B + \hbar \Omega \right) 
    + \frac{1}{2} \left(\beta_q \hbar \Omega \right)^2 B \right] ,\\
& I_{1}^{q} :=  \left(2B + \hbar \Omega \right) 
  + \epsilon \left[ \left(\beta_q \hbar \Omega \right) \left(1 + \beta_q \tilde{U}_{q}^{\mathrm{G}}\right) B 
    -  \left(\beta_q \tilde{U}_q^{\mathrm{G}}\right) \left( 1 + \frac{1}{2} \beta_q \tilde{U}_{q}^{\mathrm{G}} \right) \left(2B + \hbar \Omega \right) 
    \right] , \\
& I_{0}^{q} := B + \epsilon \left[ - \left( \beta_q \tilde{U}_{q}^{\mathrm{G}} \right) \left( 1 + \frac{1}{2} \beta_q \tilde{U}_{q}^{\mathrm{G}} \right) B \right] 
, 
\end{align}
\end{subequations}
where $A$ and $B$ are defined in Eq.~\eqref{def:A:B}.
It is noted that $\beta_q$ and $\tilde{U}_{q}$ depend on $\epsilon$. 
The function $I_{j}^{q}$ is expanded as a series of $\epsilon$ using the $\epsilon$ expansion of $\beta_q$ and of $\tilde{U}_{q}$. 
The expansion of $I_{j}^{q}$ to the order $\epsilon$ is used in the present study. 


\end{document}